# Advanced Control Strategy to Compensate Power Sharing Error and DC Circulating Current in Parallel Single-Phase Inverters

Quoc Nam Trinh
*Solar and Energy Storage Department*
*Scheneider Electric Canada*
Calgary, Alberta, Canada
trinhquocnam2010@gmail.com

Bang Nguyen
*Wind Hybrid Energy System*
*National Renewabl Energy Laboratory*
Boulder, Colorado, USA
bang.nguyen@nrel.gov

Rob Hovsapian
*ARIES Research Advisor*
*National Renewabl Energy Laboratory*
Boulder, Colorado, USA
rob.hovsapian@nrel.gov

*Abstract*—This paper proposes an advanced control strategy to eliminate both current sharing error and DC circulating current caused by line impedance mismatched and measurement errors in islanded AC microgrid system. The proposed adaptive virtual impedance scheme is developed with the aid of a low bandwidth communication and a simple PI compensator to adaptively adjust the virtual impedance value so that active and reactive power will be shared equally among inverters. Meanwhile, the DC offset in output voltage measurement is detected by using the voltage ripple information of DC link voltage of inverter. The proposed control strategy can be implemented directly without any pre-knowledge of the feeder impedances. The proposed compensation scheme is simple and easy to implement, and it also does not require extra hardware circuit or sensors. The effectiveness of the proposed solution is verified by simulation and experimental results of 6 kVA microgrid system.

## I. INTRODUCTION

The utilization of renewable energy sources in the form of distributed generation (DG) units has been greatly increased in recent decades to fulfil the increasing demand for electricity and reduce stress on the existing transmission system [1]. With an increasing numbers of DG units integrated into existing power system, the microgrid concept has been emerged as a feasible and effective approach to achieve better operation of multiple DG units [2]. A microgrid should able to flexibly operate in either a grid-connected or an islanding mode to improve the reliability and quality of power supplied to customers [3], [4].

When the microgrid operates in the islanding mode, each DG unit must supply active and reactive power and share the total load in proportion to its power rating to prevent the overload condition for each DG unit. In order to perform the power sharing requirement with a decentralized manner, the conventional real power-frequency and reactive power-voltage magnitude droop control method have been adopted [4]-[6]. The major advantages of droop control methods are simple to implement, only use local measurement, and does not require any communication between DG units. However, due to the mismatched line impedances and different local loads, the reactive power sharing among DG units is significantly affected and normally not possible to achieve equal reactive power sharing [7]. In addition, when the local loads are unbalance or nonlinear loads, harmonic power sharing accuracy cannot be ensured due to mismatched grid impedances [8].

In order to improve power sharing accuracy among DG units, a few improved droop control methods have been introduced by introducing virtual impedance control methods [9]-[15]. With this kind of control methods, the inverter output impedance is modified to eliminate the mismatch among inverter equivalent impedances. The virtual impedances are normally generated with proportional-integral controller (PI) regulators according to reactive power [12] or active power error [13] or current error [14]. In addition, harmonic power sharing strategies were also investigated by utilizing harmonic virtual impedances [15].

Most of previous studies only investigate the impact of mismatch line impedance on power sharing performance. However, there is another important factor that may affect performance of parallel inverters: DC offset and scaling errors of voltage and current measurements [16]. Impact of measurement errors on control performance of inverters or AC/DC converters has been studied in [17]-[19]. Some compensation schemes are also proposed to mitigate the negative impact of measurement errors [18]-[21]. However, in microgrid system, since multiple inverters are operating in parallel and they only can measure local voltage and current, it become more challenge to detect and compensate sensor measurement errors. Compensation of measurement errors for microgrid system has been introduced in [16] by combining feed-back and feed-forward voltage control. However, this paper only focuses on compensating the measurement error without considering impact of mismatch line impedance on the inverter performance. In addition, the system under investigation is three phase system, which is different from single phase system presented in this paper.

This paper proposes an advanced control strategy to deal with both voltage measurement error and line impedance mismatch to compensate both alternative current (AC) and direct current (DC) circulating current in parallel inverters. The line impedance mismatch is compensated by an adaptive virtual impedance scheme. The proposed AVI is developed with the aid of a low bandwidth communication (LBC) and a simple PI compensator to adaptively adjust the virtual impedance value so that active and reactive power will be shared equally among inverters regardless of physical impedance mismatch. Meanwhile, the voltage measurement is detected by utilizing voltage ripple information at DC link side of inverter. The proposed control strategy can be implemented directly without any pre-knowledge of the feeder impedances. The proposed compensation scheme is simple and easy to implement, and it also does not require extra hardware circuit or sensors. The effectiveness of the proposed solution is verified by experimental results of 6 kVA microgrid system.

II. SYSTEM AND OPERATION OF ISLANDED MICROGRID

## A. Conventional Droop Control method for Single phase DG unit

Fig. 1 shows the configuration of a islanded AC microgrid system, where the microgrid is composed of a number of DG units and loads. Each DG unit is supplied by a DC-link, inverter, and LC filter; the inverters are controlled by the local controllers. The Microgrid central controller (MGCC) monitors the microgrid and main grid status to define whether the microgrid is operating in grid-connected mode or islanded mode.

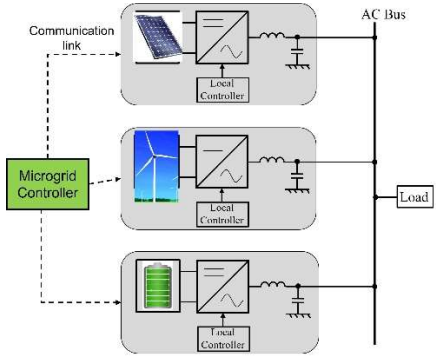

Fig. 1 System configuration of islanded AC microgrid.

In this paper, the microgrid system under investigation is mainly resistive line feeder impedance. Therefore, the magnitude and angular frequency of the microgrid system is described as follows

$$V = V_0 - mP_{LPF}$$
$$\omega = \omega_0 + nQ_{LPF} \quad (1)$$

where $\omega_0$ and $V_0$ are the nominal values of the DG angular frequency and DG voltage magnitude, respectively; P and Q are the measured real and reactive powers after the low-pass filter (LPF), respectively; and m and n are the real and reactive power droop slopes, respectively.

Fig. 3 illustrates the basic control scheme for DG unit with droop control function described in (2). From droop control function, the reference voltage $v_{ref} = V\sin(\omega t)$ is generated to provide reference voltage of each DG unit. The voltage control scheme of each DG unit is consisted of dual voltage and current control loops where both controllers are implemented using Proportional plus three resonant controllers tuned at the fundamental frequency ($\omega_s$), third order and fifth order harmonics. The transfer function of P-3R controller is described as follows:

$$G_v(s) = K_p + \sum_{1,3,5} \frac{2K_{rh}s}{s^2 + (h\omega_s)^2} \quad (2)$$

Where $K_p$ and $K_{rh}$ are proportional and resonant gains of P-3R controller. Assuming that output voltage and current of each DG units contain of AC components and DC offsets caused by ADC offset measurement errors as follows:

$$v_{inv1}(t) = V_{m1}\cos(\omega t) + V_{dc1} + V_{adc-offset1} \quad (3)$$
$$v_{inv2}(t) = V_{m2}\cos(\omega t) + V_{dc2} + V_{adc-offset2}$$

where $V_{adc-offset}$ and $V_{adc-offset}$ are ADC offset error of the inverter output voltage measurement. The line feeder impedance at output of each DG unit are Z1 and Z2. The circulating DC and AC current is calculated as:

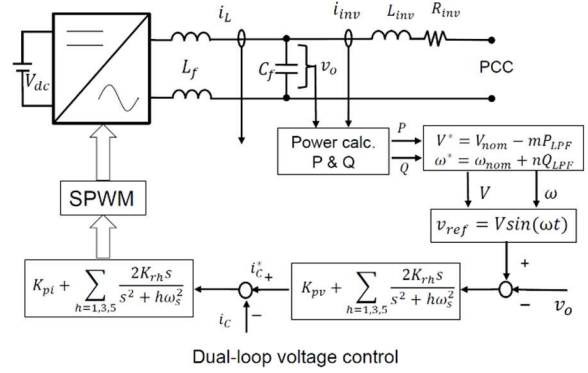

Fig. 2. Basic control scheme for DG unit with droop control.

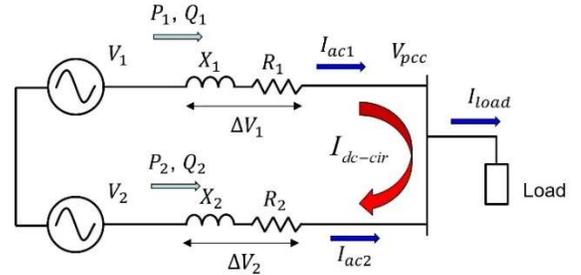

Fig. 3. Current sharing error and DC circulating current in parallel inverters.

$$\Delta I = \frac{(V_{m1}\cos(\omega t) + V_{adc-offset})}{Z_1}$$
$$- \frac{(V_{m2}\cos(\omega t) + V_{adc-offset})}{Z_2} \quad (4)$$

Where the ciculating DC current and AC current sharing error are determined as follows:

$$I_{dc-cir} = \frac{V_{dc1} + V_{adc-offset}}{Z_1} - \frac{V_{dc2} + V_{adc-offset}}{Z_2}$$
$$\Delta I_{ac} = \frac{V_{m1}\cos(\omega t)}{Z_1} - \frac{V_{m2}\cos(\omega t)}{Z_2} \quad (5)$$

From (5), it can be seen that the DC circulating current is caused by DC offset error and line impedance difference of two DG units. Similarly, even we can control to generate exactly same inverter output voltage, the current sharing error $\Delta I_{ac}$ is also generated by the line impedance mismatch $Z_1 \neq Z_2$ [14].

The impact of ADC offset error and the procedure to detect measurement errors are described in following section.

*B. DC circulating current problem caused by ADC offset measurement errors*

Instantaneous output power of single-phase inverter and concept to detect DC offset voltage and current measurement errors.

$$v_{inv}(t) = V_m \cos(\omega t) + V_{adc-offse} \quad (6)$$

$$i_{inv}(t) = I_m \cos(\omega t - \varphi) + I_{dc-cir} \quad (7)$$

Instantaneous output power of inverters described as follows:

$$\begin{aligned}p_{out}(t) &= v_{inv}(t) * i_{inv}(t) \\ &= \left(V_m \cos(\omega t) + V_{adc-offset}\right) * (I_m \cos(\omega t - \varphi) + I_{dc-cir}) \\ &= V_m I_m \cos(\omega t)\cos(\omega t - \varphi) + V_m I_{adc-offset}\cos(\omega t) \\ &\quad + V_{adc-offset} I_m \cos(\omega t - \varphi) + V_{adc-offset} I_{dc-cir} \\ &= \tfrac{1}{2} V_m I_m \cos(\varphi) + \tfrac{1}{2} V_m I_m \cos(2\omega t - \varphi) + V_m I_{DC}\cos(\omega t) \\ &\quad + V_{dc} I_m \cos(\omega t - \varphi) + I_{dc-cir} V_{adc-offset} \\ &= p_{out-dc}(t) + p_{out}(t) + p_{out-2\omega}(t) \end{aligned} \quad (8)$$

From (8), we can see that instantaneous output power of inverter is composed of three components, DC power, pulsation power at fundamental ripple and double frequency ripple as follows:

$$p_{out-dc}(t) = \tfrac{1}{2} V_m I_m \cos(\varphi) + V_{adc-offset} I_{dc} \quad (9)$$

$$p_{out-\omega}(t) = V_m I_{dc-cir}\cos(\omega t) + V_{adc-offset} I_m \cos(\omega t - \varphi)$$

$$p_{out-2\omega}(t) = \tfrac{1}{2} V_m I_m \cos(2\omega t - \varphi)$$

Assuming that the power loss of inverter is very small and negligible, the ripple power $p_{out-\omega}$ in (8) will also reflect to the ripple on the DC-link voltage of inverter.

$$p_{in} = v_{dc} * i_{dc} \approx p_{out} = p_{out-dc} + p_{out-\omega} + p_{out-2} \quad (10)$$

As a result, the ADC offset errors in voltage measurements will result in power ripple at fundamental frequency on the DC link voltage of each inverter and degrade performance of output current. We can utilize this information to detect and compensate the ADC offset error in the inverter unit.

## III. PROPOSED CONTROL STRATEGY TO COMPENSATE CURRENT SHARING ERROR AND DC OFFSET MEASUREMENT ERROR

The overall control scheme of the proposed method is illustrated in Fig. 4 with adaptive virtual impedance (AVI) control to minimize current sharing error and the voltage offset compensation is introduced to remove the DC offset in voltage measurement error. Details of compensation method and AVI scheme are explained in following sub-sections.

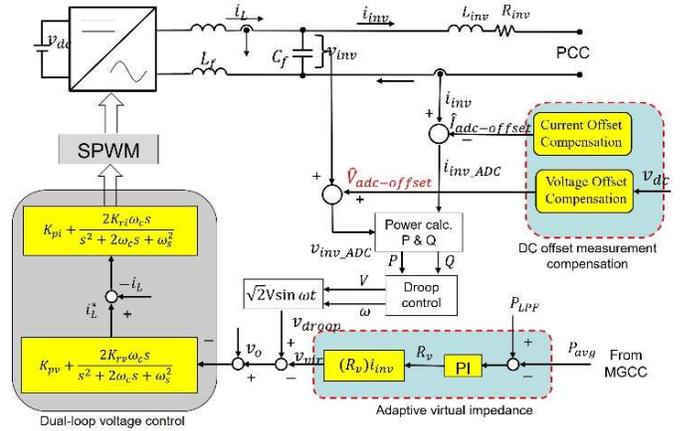

*Fig. 4.* Proposed AVI and Measurement error compensation strategy.

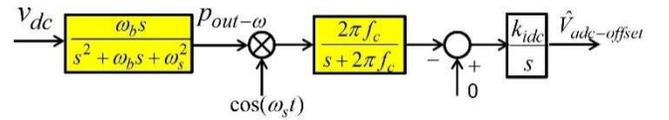

*Fig. 5.* Concept to detect and compensate for DC offset error in Voltage measurement.

*A. Proposed DC Offset Compensation Method for Voltage Measurement Errors*

From (8) and (9), the ADC offset of voltage measurement $V_{adc-offset}$ can be detected from output power ripple at $\omega_s$ from instantaneous output power of the inverter, which is the ripple power $p_{out-\omega}$ in (8). However, it is not possible to measure $p_{out-\omega}$, we can based on the voltage ripple at DC-link side of the inverter to determine the $p_{out-\omega}$.

From (10), the ripple power $p_{out-\omega}$ in (8) will also reflect to the ripple on the DC-link voltage of inverter. Therefore, we can detect $p_{out-\omega}$ indirectly through applying a band-pass filter (BPF) through the ripple voltage at $v_{DC}$ as follows:

$$p_{ac-\omega} = \text{BPF}\{p_{out}\} = \frac{\omega_b s}{s^2+\omega_b s+\omega^2}\{v_{DC}\} \quad (11)$$

After getting the $p_{ac-\omega}$, multiplying the output with $\cos(\omega t)$ function and we can estimate $V_{adc-offse}$ in (12) by applying a low-pass filter (LPF) to (12) and we achieve a pure DC component of $V_{adc-offset}$ in (13).

$$\begin{aligned}p_{ac-\omega}\cos(\omega t) &= V_m I_{adc-offset}\cos(\omega t)\cos(\omega t) \\ &= \tfrac{1}{2} V_m I_{adc-offset}(1 + \cos(2\omega t)) \end{aligned} \quad (12)$$

$$V_{adc-offset} = \text{LPF}\{p_{ac-\omega}\cos(\omega t)\}/\tfrac{1}{2}V_m \quad (13)$$

The operational concept of the ADC offset detection in current measurement is summarized in Fig. 5.

*B. Proposed Adaptive Virtual Impedance Control*

The adaptive virtual impedance control diagram is shown in Fig. 4. The inverter transmits the information of the respective active power outputs (P1, P2,…, and Pn) to the MGCC. The MGCC determines the total active power supplied by the inverters in the microgrid by considering the total rated active power of the inverters, and average active power of whole microgrid system is calculated as

$$P_{avg} = \frac{\sum P_i}{N} \quad (14)$$

The MGCC then sends this average power data to all DG units by mean of a low band-width communication network. From the information of average power and local output power of each DG unit, the virtual impedance of each DG unit is estimated through as Proportional Integral (PI) compensator as follows:

$$\hat{R}_v = (K_p + \frac{K_i}{s})(P_{avg} - P_{DGi}) \quad (15)$$

Where $K_p$ and $K_i$ are controller gains of PI compensator for adaptive virtual impedance compensation.

After getting the virtual impedance value $R_v$, this virtual impedance will multiply with instantaneous output current value to estimate voltage drop and compensate to the voltage reference generation as follows:

$$v_{vir} = (\hat{R}_v)i_{inv} \quad (16)$$

## IV. SIMULATION AND EXPERIMENTAL RESULTS

To validate the effectiveness of the proposed control method, simulation and experimental tests of parallel inverters were carried out with two parallel inverters as shown in Fig. 6. The proposed control strategy is implemented using TI C2000 DSP TMS320F28379D. The switching and control frequency of PWM signal is set at 20 kHz. Low Bandwidth Communication is implemented using CAN bus communication at the baud rate of 500kps. The main parameters of hardware system are listed in Table I and Table II.

A simulation test of two inverters operating in parallel with the line impedance mismatched and voltage measurement offset error provided in Table II. The simulation results are shown in Fig. 7. In Fig.7, it can be seen that the output current of inverter 1 and 2 are not balanced and contain DC offset. The circulating current also contain DC offset with amplitude of -5.5A and ac ripple at 2.5A RMS. This result validates the impact of mismatch line impedance and DC offset in voltage measurement of inverter output and the conventional droop control is not able to mitigate the impact of those factors on current sharing.

The same simulation is conducted using the proposed control strategy with AVI and measurement compensation scheme and the results are presented in Fig. 8. In Fig. 8, the inverter output current is balanced with the same amplitude, where we can see the two currents are overlap. In addition, the DC offset in circulating current has been eliminated, only small AC current ripple exist with the amplitude of 0.7A, which is much smaller than the results shown in Fig. 7. These results verify the effectiveness of the proposed control method.

**Table I**. Hardware system Parameters

| Parameters | Values | Parameters | Values |
|---|---|---|---|
| Nominal output voltage | 200 V | Switching frequency ($f_{sw}$) | 20 kHz |
| Nominal output frequency | 50 Hz | Rload | 14 Ohm |
| Rated output power of one inverter | 2 kVA | Droop frequency coefficient (m) | 0.001 |
| DC-link voltage ($V_{dc}$) | 250 V | Droop voltage coefficient (n) | 0.0025 |
| Output filter ($L_f$) | 0.5 mH | $V_{adc-offset1}$ | -5.0V |
| Output filter capacitance ($C_f$) | 15 μF | $V_{adc-offset2}$ | 0V |

**Table II**. Parameters of Line impedance of DG units

| Balanced Impedance | Mismatched Impedance |
|---|---|
| R1 = 0.22Ohm | R1 = 0.44Ohm |
| R2 = 0.22Ohm | R2 = 0.22Ohm |

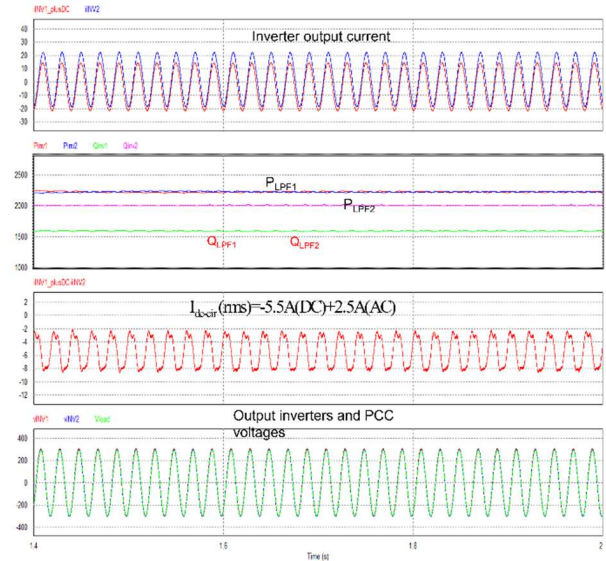

*Fig. 7.* Simulation results of two inverters with only conventional droop control.

Fig. 9(a) shows experimental results of two inverters in parallel to supply a linear load without adaptive virtual impedance. As shown in Fig. 6(a), the two inverters have very poor current sharing, the shape of each inverter output current is not sinusoidal and the circulating current between two inverters contain DC offset and AC component with the amplitude of

2.44A This occurs due to the mismatch of line feeder impedance at output of each inverter and DC offset in the voltage measurement. The inverter control scheme is not able to detect the mismatch line impedance and perform necessary compensation.

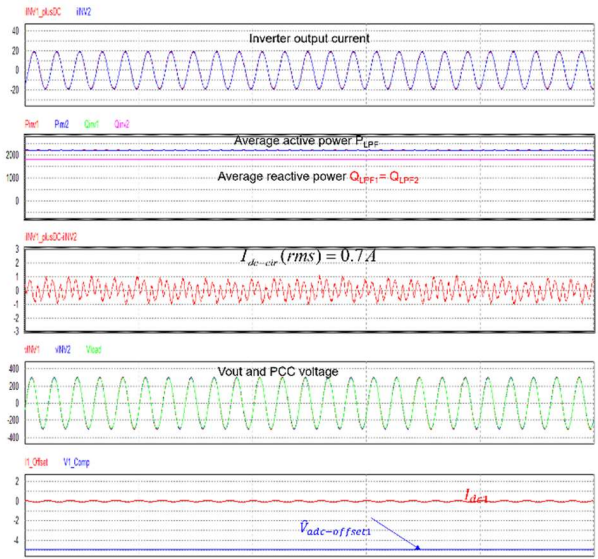

*Fig. 8*. Simulation results of two inverters with proposed control method.

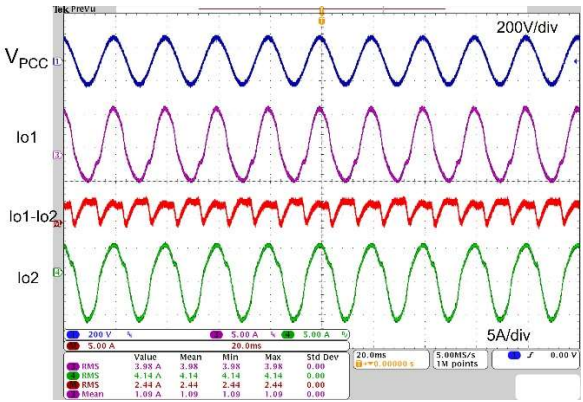

(a) With conventional droop control

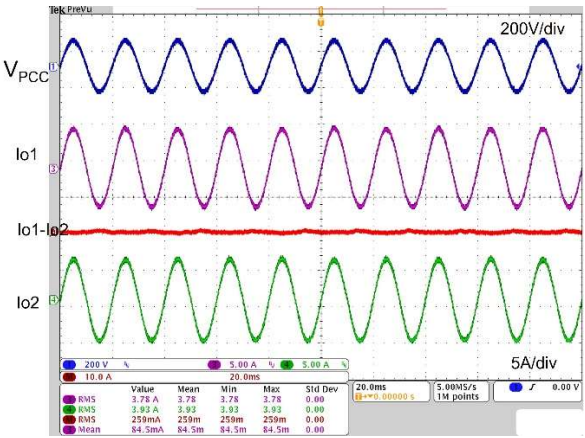

(b) With proposed compensation method

*Fig. 9* Experimental results of two inverters in parallel to supply a linear load without adaptive virtual impedance.

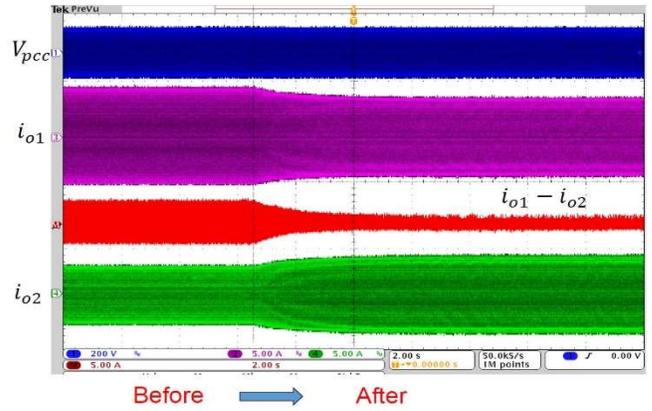

(a)

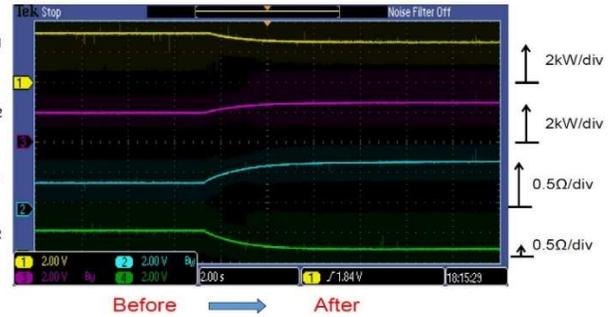

(b)

*Fig. 10* Dynamic test when activate the AVI method when system running.

In order to improve that issue, the same test case is performed with the proposed control method and the results are shown in Fig. 9(b). In Fig. 9(b), the output current of each inverter is sinusoidal and has the same amplitude. The circulating current between the two inverters are significantly reduced to 0.25A, which is 10 times smaller than which is shown in Fig. 9(a). In addition, no DC offset exist in the circulating current. This experimental result successfully verified the effectiveness of the proposed compensation method.

Fig. 10 shows the experimental test when the proposed AVI method is activated on-line while system in operation. Before the AVI method is activated, the two inverters supply different output current and the circulating current is significant. Once the AVI is on active, the first inverter current ($i_{o1}$) reduce the amplitude and the second inverter current ($i_{o2}$) increase amplitude to balance the output power. As a result, the circulating current ($i_{o1} - i_{o2}$) is also reduced significantly.

Fig. 10(b) shows the output Rv1 and Rv2 of the AVI method on each inverter and the output power of each inverter. Once the compensating method is ON, the output power of each inverter is balanced. This experimental result proved that the proposed AVI method does not affect the reliability and stability of system.

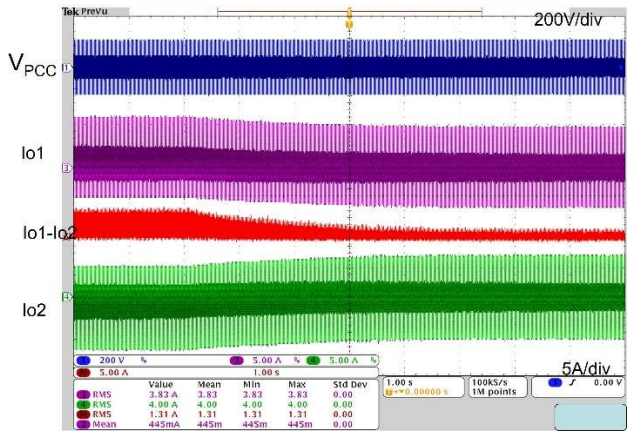

*Fig. 11.* Dynamic response when activate DC offset compensation.

Fig. 11 shows the experimental test when the proposed DC offset compensation method is activated while system is in operation. Before the DC offset compensation scheme is activated, the two inverters have a different output current, and the circulating current contain both AC component and DC offset. Once the AVI is on active, the circulating current ($i_{o1} - i_{o2}$) is also reduced significantly. This result validated the effectiveness of the proposed DC offset compensation method without compromising system stability.

## V. CONCLUSIONS

This paper proposed an advanced control method to compensate impact of measurement errors and line impedance mismatched on current sharing performance of parallel inverters. By adjusting the virtual impedances at the outputs of DG units, the accurate current and active power sharing is achieved even though the line impedance mismatches with each other. Meanwhile the DC offset in voltage measurement is detected and compensated to remove DC circulating current. The proposed control strategy does not require any feeder impedance, which eliminates the need of online feeder impedance estimator and decreases the system complexity. The effectiveness of the proposed method is verified with simulation and experimental tests.